\documentclass[aps,pra,preprint,superscriptaddress,longbibliography]{revtex4-2}

\usepackage{amsmath}
\usepackage{amsthm}
\usepackage{amsfonts}
\usepackage{amssymb}
\usepackage[dvipsnames]{xcolor}
\usepackage{graphicx}
\usepackage{color}
\usepackage[pdftex]{hyperref} 
\usepackage{longtable}
\usepackage{array}
\usepackage{dsfont}
\usepackage{bm}
\usepackage{braket}

\begin{document}

\title{Non-Hermitian $\mathrm{sl}(3, \mathbb{C})$ three-mode couplers}

\author{B.~M. Rodriguez-Lara}
\affiliation{Universidad Polit\'ecnica Metropolitana de Hidalgo, Tolcayuca, Hidalgo 43860, Mexico.}
\email{blas.rodriguez@gmail.com}

\author{H. Ghaemi-Dizicheh}
\affiliation{Department of Physics and Astronomy, University of Texas Rio Grande Valley, Texas 78539, USA.}

\author{S. Dehdashti} 
\affiliation{Emmy-Noether Group Theoretical Quantum Systems Design, Technical University of Munich, Munich, Germany.}

\author{A. Hanke}
\affiliation{Department of Physics and Astronomy, University of Texas Rio Grande Valley, Texas 78539, USA.}

\author{A. Touhami}
\affiliation{Department of Physics and Astronomy, University of Texas Rio Grande Valley, Texas 78539, USA.}

\author{J. N\"{o}tzel} 
\affiliation{Emmy-Noether Group Theoretical Quantum Systems Design, Technical University of Munich, Munich, Germany.}

\begin{abstract}
Photonic systems with exceptional points, where eigenvalues and corresponding
eigenstates coalesce, have attracted interest due to their topological features and enhanced sensitivity to external perturbations.
Non-Hermitian mode-coupling matrices provide a tractable analytic framework to model gain, loss, and chirality across optical, electronic, and mechanical platforms without the complexity of full open-system dynamics.  
Exceptional points define their spectral topology, and enable applications in mode control, amplification, and sensing.
Yet $N$-mode couplers, the minimal setting for $N$th-order exceptional points, are often studied in specific designs that overlook their algebraic structure.
We introduce a general $\mathrm{sl}(N,\mathbb{C})$ framework for arbitrary $N$-mode couplers in classical and quantum regimes, and develop it explicitly for $N=3$. 
This case admits algebraic diagonalization, where a propagation-dependent gauge aligns local and dynamical spectra and reveals the geometric phase connecting adiabatic and exact propagation.
An exact Wei--Norman propagator captures the full dynamics and makes crossing exceptional points explicit.
Our framework enables classification of coupler families. 
We study the family spanning $\mathcal{PT}$-symmetric and non-Hermitian cyclic couplers, where two exceptional points of order three lie within a continuum of exceptional points of order two, ruling out pure encircling.
As an application, we study these exceptional points for a lossy three-leg beam splitter and reveal its propagation dynamics as a function of initial states, such as Fock and NOON states.
Our approach provides a systematic route to analyze non-Hermitian mode couplers and guide design in classical and quantum platforms.
\end{abstract}

\maketitle


\section{Introduction}
Photonic systems provide versatile platforms to explore non-Hermitian physics across classical and quantum regimes. 
Coupled waveguides emulate tight-binding lattices \cite{Lin2025p223802,Keil2016p213901,Chien2007p125113,Chen2021p023501}, simulate condensed-matter phenomena \cite{Yang2024p50,Wang2024p063525,Jorg2025pPC133930D}, and realize controllable settings for engineered gain, loss, and symmetry constraints \cite{Xiao2025pR09,Li2024p156601,Reisenbauer2024p1629}. 
Within this context, exceptional points (EPs), corresponding to parameter values where eigenvalues and eigenvectors coalesce \cite{Kato1966}, constitute central spectral features of non-Hermitian systems, giving rise to nontrivial topology and enhanced susceptibility to perturbations in optics and photonics \cite{Miri2019p7709}.
Recent pseudospectral analyses show that sensitivity near exceptional points depends on the symmetry class of accessible perturbations \cite{Komis2022p043219}. 
Moreover, information-theoretic analyses show that apparent enhancements in non-Hermitian sensing must be assessed in terms of effective metrological resources once post-selection and environmental degrees of freedom are included \cite{Zeng2025}.

In photonic implementations, the interplay of coherent coupling, non-Hermiticity, and parity--time ($\mathcal{PT}$) symmetry \cite{Bender1998p5243,Bender1999p2201,Bender2002p270401} provides a concrete setting in which these phenomena arise, enabling symmetry-breaking transitions and exceptional points in coupled lasers \cite{Yao2016p}, periodic $\mathcal{PT}$-modulated waveguides \cite{Zhong2022p52504}, and slab geometries \cite{Nolting1996p76,ElGanainy2007p2632,Zhang2024p053515}.
In such systems, generalized coupled-mode theory offers a natural description of non-Hermitian waveguides and resonators, capturing mode hybridization, gain/loss compensation, and non-reciprocal light transport \cite{Zhang2015p22653,Ghaemidizicheh2025p26329}.

Open quantum systems are rigorously described by Lindblad or quantum Langevin formalisms, which are often analytically intractable \cite{Bergholtz2021p015005}. 
Non-Hermitian mode-coupling matrices instead provide minimal effective models for open systems with gain, loss, or chirality \cite{Ashida2020p249}. 
Such effective descriptions arise in two controlled regimes. 
First, as reductions of Lindblad or Langevin dynamics at the mean-field or single-excitation level. 
Second, as generators of post-selected evolution in open quantum systems, where the resulting dynamics is confined to fixed-excitation subspaces.
Within this scope, non-Hermitian mode-coupling models capture complex eigenvalues, non-orthogonal eigenmodes, and eigenvalue–eigenvector coalescence at exceptional points \cite{JaramilloAvila2020p1761,QuirozJuarez2019p862,Ding2022p745}. 
These effective models apply across diverse platforms, including optical waveguides and resonators \cite{Dehdashti2015p067132,Setare2019p1663,Wang2023p442}, electronic systems \cite{QuirozJuarez2022p054034}, transport effects \cite{GhaemiDizicheh2021p023515,GhaemiDizicheh2023p125155}, nonlinear skin effects \cite{GhaemiDizicheh2024p125411}, superconducting circuits \cite{Song2024p9848}, acoustic metamaterials \cite{Achilleos2017p144303}, mechanical structures \cite{Lu2023p202307998}, and ultracold atomic systems \cite{Wang2022p4598}.

One approach to organizing the spectral structure of non-Hermitian couplers is graph-theoretical, encoding coupler architectures as directed graphs whose adjacency matrices capture spectral topology \cite{Tschernig2020p1161,Tschernig2022p2100707,Grom2025p23132}. 
For discrete photonic platforms, this adjacency matrix is proportional to the mode-coupling matrix. 
Here we adopt a symmetry-guided framework that treats the full class of mode-coupling matrices with complex elements at the level of their generating algebra, rather than individual realizations.

Exceptional points are critical branch points of the eigenvalue surfaces where eigenvalues and eigenvectors coalesce \cite{Kato1966,Heiss2012p444016}. 
Their topological nature manifests in the response to parameter-space evolution.
Closed paths that do not enclose an EP yield cyclic evolution, with the system returning to its initial state up to a geometric phase. 
In contrast, paths that enclose an EP of order $m$ induce a nontrivial permutation of eigenstates, requiring $m$ encirclements to return to the initial state.
Second-order exceptional points (EP2s) have been realized in coupled waveguides with engineered gain and loss \cite{Ruter2010p192,Regensburger2012p167}, microring resonators \cite{Peng2014p394}, and photonic crystal slabs \cite{Zhen2015p354}. 
At these degeneracies, systems exhibit non-reciprocal transport \cite{Feng2013p108,Laha2024p033511}, loss-induced transparency \cite{Guo2009p093902}, and enhanced sensitivity with a square-root scaling of perturbations \cite{Wiersig2014p203901}. 
These effects enable applications from low-threshold lasing \cite{Zamir2023p063504} to sensing beyond the quantum limit \cite{Miri2019p7709,Zhang2025p344,Lambert2025p}, and motivate the exploration of higher-order EPs. 
At an $m$th-order EP, the eigenvalue splitting follows an $m$th-root dependence on perturbations \cite{Heiss2008p244010,Heiss2016p495303,Demange2012p025303,Ryu2012p042101,Shi2025pL022034}. 
This scaling motivates their use in precision metrology and weak-field detection. 
Realizing them remains challenging, as enforcing higher-order degeneracies requires fine-tuning multiple parameters \cite{Teimourpour2018p261}, but waveguide-based quantum beam splitters offer a concrete route to overcome this constraint \cite{QuirozJuarez2019p862,Ghaemidizicheh2025p26329}.

Non-Hermitian three-mode couplers provide the minimal setting for third-order exceptional points (EP3s). 
They have been demonstrated in $\mathcal{PT}$-symmetric waveguides \cite{Schnabel2017p053868,ZaragozaGutierrez2016p3989} and microcavities \cite{Laha2020p063829,Kullig2023pA54}, yet most analyses remain tied to specific architectures, overlooking the underlying algebraic structure that governs their spectral topology and dynamics. 
Here, we develop a symmetry-guided framework to expose the general features of non-Hermitian $3\times3$ couplers and provide a systematic route to their classification.
In Sec.~\ref{sec:Sec2}, we construct a general framework for non-Hermitian mode-coupling matrices as elements of the Lie algebra $\mathrm{gl}(N,\mathbb{C})$. 
A propagation-dependent scalar gauge removes the trace, leaving traceless matrices in $\mathrm{sl}(N,\mathbb{C})$. 
We extend the framework to the quantum regime using an $N$-boson bilinear representation that conserves the excitation number within the effective non-Hermitian generator and realizes irreducible $\mathrm{sl}(N,\mathbb{C})$ multiplets. 
In Sec.~\ref{subsec:effectivecoupler}, we apply the formalism to the three-mode coupler in the isospin--hypercharge representation of $\mathrm{sl}(3,\mathbb{C})$. 
Cartan generators encode relative phase and differential gain, while ladder generators describe coherent mode coupling. 
We analyze the spectral structure through local and dynamical gauges in Sec.~\ref{subsec:spectralanalysis}. 
Exact propagators are constructed via a Wei--Norman factorization in Sec.~\ref{subsec:propdyn}, where we identify the role of holonomy in distinguishing adiabatic and dynamical evolution. 
Higher-order exceptional points are analyzed in Sec.~\ref{subsec:hoeps}. 
As an application, in Sec.~\ref{sec:PTsymcc} we explore $\mathcal{PT}$-symmetric and non-Hermitian cyclic couplers, where third-order EPs are embedded in continua of second-order EPs, illustrating the complexity of their spectral landscape and motivating symmetry-based design principles for photonic devices. 
We conclude in Sec.~\ref{sec:conclusions}.

\section{Optical and Photonic Models} \label{sec:Sec2}

We begin with the classical mode-coupling model to establish the minimal framework where the algebraic structure of non-Hermitian systems with gain, loss, and directional coupling is transparent.
In this framework, the complex field amplitudes propagate under an effective Schr\"odinger-like equation,
\begin{align}
    -i \, \partial_{z} \bm{E}(z) = \bm{M}(z)  \bm{E}(z),
\end{align}
where $z$ is the longitudinal propagation coordinate, $\bm{E}(z)$ is the vector of mode amplitudes $E_{j}(z)$, and $\bm{M}(z)$ is a mode-coupling matrix with dimensions of spatial frequency and elements
\begin{align}
    \left[ \bm{M}(z) \right]_{j,k} = a_{j,k}(z) + i b_{j,k}(z),
\end{align}
with $a_{j,k}(z), b_{j,k}(z) \in \mathbb{R}$. 
Diagonal terms $a_{j,j}(z)$ represent the effective propagation constants, while $b_{j,j}(z)$ describe local gain or loss. 
Off-diagonal terms $a_{j,k}(z)$ and $b_{j,k}(z)$ for $j \neq k$ define the directional coupling between modes. 

The mode-coupling matrix $\bm{M}(z)$ belongs to the Lie algebra $\mathrm{gl}(N,\mathbb{C})$ of all complex $N \times N$ matrices. 
This algebra admits the standard basis $\bm{O}_{j,k}$ with a one at position $(j,k)$ and zeros elsewhere, satisfying 
\begin{align}
    \left[ \bm{O}_{j,k}, \bm{O}_{m,n} \right] = \delta_{k,m} \bm{O}_{j,n} - \delta_{j,n} \bm{O}_{m,k}.
\end{align}
We introduce a gauge redefinition on the field amplitudes,
\begin{align}
    \bm{E}(z) =
    e^{ \frac{i}{N} \int_{0}^{z} \mathrm{d}\zeta \, \mathrm{Tr} \left[ \bm{M}(\zeta) \right] } \bm{E}_{1}(z),
\end{align}
to remove global phase and uniform gain or loss. 
The transformed coupling matrix,
\begin{align}
    \bm{M}_{1}(z) = \bm{M}(z) - \frac{1}{N} \mathrm{Tr} \left[ \bm{M}(z) \right] \bm{1},
\end{align}
belongs to the subalgebra $\mathrm{sl}(N,\mathbb{C}) \subset \mathrm{gl}(N,\mathbb{C})$ of traceless complex matrices.
Its characteristic polynomial 
\begin{align}
    p(\lambda) = \sum_{k=0}^{N} \beta_{k}(z) \lambda^{N-k},
\end{align}
has coefficients $\beta_{0}(z) = 1$, $\beta_{1}(z) = -\mathrm{Tr} \left[ \bm{M}_{1}(z) \right] = 0$, and 
\begin{align}
    \beta_{k}(z) = - \frac{1}{k} \sum_{j=1}^{k} \beta_{k-j}(z) \mathrm{Tr} \left[ \bm{M}_{1}^{j}(z) \right], \qquad k \ge 1,
\end{align}
generated by trace invariants \cite{Hou1998p706}.
Spectral degeneracies correspond to eigenvalue coalescence in the non-Hermitian case and occur when the discriminant vanishes \cite{Cox1997},
\begin{align}
    \Delta(p)=(-1)^{\tfrac{N(N-1)}{2}}\,\mathrm{Res}\left(p,p'\right),
\end{align}
where $\mathrm{Res}(p,p')$ is the resultant of $p(\lambda)$ and its derivative, defined as the determinant of their Sylvester matrix. 
The resultant vanishes exactly when $p$ and $p'$ share a root, that is, when $p$ has a repeated eigenvalue.
In this traceless representation, any $N$th-order exceptional point collapses to the zero eigenvalue, since $\mathrm{Tr} \, \left[ \bm{M}_{1}(z) \right] = N \lambda_{0} = 0$.
Exceptional points of lower order may occur at zero or at non-zero eigenvalues.

We extend the classical description into a photonic model \cite{RodriguezLara2011p053845,MunozEspinosa2022p10505} by promoting optical mode amplitudes to Fock states,
\begin{align}
    \left\vert E_{j}(z) \right\rangle = \sum_{k=0}^{\infty} c_{k}^{(j)} \left\vert k \right\rangle,
\end{align}
with $c_{k}^{(j)} \in \mathbb{C}$ and $\left\vert k \right\rangle$ the $k$-photon state. 
Normalization does not hold in general because non-Hermitian dynamics allow amplification or decay. 
We promote the mode-coupling to the $N$-boson representation,
\begin{align}
    \hat{M}(z) =  \sum_{j,k=1}^{N} \left[ \bm{M}(z) \right]_{j,k} \, \hat{a}^{\dagger}_{j} \hat{a}_{k},
\end{align}
with bosonic ladder operators,
\begin{align}
    \left[ \hat{a}_{j}, \hat{a}^{\dagger}_{k} \right] = \delta_{j,k}.
\end{align}
The operator basis,
\begin{align}
    \hat{O}_{j,k} = \sum_{m,n=1}^{N} \left[ \bm{O}_{j,k} \right]_{m,n} \hat{a}^{\dagger}_{m} \hat{a}_{n},
\end{align}
satisfies the same commutation relations,
\begin{align}
    \left[ \hat{O}_{j,k}, \hat{O}_{m,n} \right] = \delta_{k,m} \hat{O}_{j,n} - \delta_{j,n} \hat{O}_{m,k}.
\end{align}
Thus $\hat{O}_{j,k}$ realize $\mathrm{gl}(N,\mathbb{C})$ on Fock space, and restriction to fixed excitation number yields finite-dimensional representations of $\mathrm{sl}(N,\mathbb{C})$.

We again remove the shared phase and amplification by a gauge redefinition,
\begin{align}
    \left\vert E(z) \right\rangle = 
    e^{ \frac{i}{N} \int_{0}^{z} \mathrm{d}\zeta \, \mathrm{Tr} \left[ \bm{M}(\zeta) \right] }
    \left\vert E_{1}(z) \right\rangle,
\end{align}
yielding the effective mode-coupling traceless operator
\begin{align}
    \hat{M}_{1}(z) = \hat{M}(z) - \frac{1}{N} \mathrm{Tr} \left[ \bm{M}(z) \right] \hat{n},
\end{align}
with $\hat{n} = \sum_{j=1}^{N} \hat{a}^{\dagger}_{j} \hat{a}_{j} \equiv \sum_{j=1}^{N} \hat{n}_{j}$ the total excitation number operator.
Thus both classical and photonic models share the same algebraic structure, realizing $\mathrm{sl}(N,\mathbb{C})$.

\subsection{Effective \texorpdfstring{$\mathrm{sl}(3,\mathbb{C})$}{sl(3,C)} coupler} 
\label{subsec:effectivecoupler}
We focus on the effective non-Hermitian three-mode coupler in the traceless similarity frame
\begin{align}
    \bm{M}_{1}(z) =
    \begin{pmatrix}
        \mu_{1,1}(z) & \mu_{1,2}(z) & \mu_{1,3}(z) \\
        \mu_{2,1}(z) & \mu_{2,2}(z) & \mu_{2,3}(z) \\
        \mu_{3,1}(z) & \mu_{3,2}(z) & \mu_{3,3}(z)
    \end{pmatrix},
\end{align}
with $\mu_{j,k}(z) = a_{j,k}(z) + i b_{j,k}(z)$ and diagonal terms $\mu_{j,j}(z) = \frac{2}{3} \left[ a_{j,j}(z) + i b_{j,j}(z) \right] - \frac{1}{3}\left[  a_{k,k}(z) + i b_{k,k}(z) +  a_{l,l}(z) + i b_{l,l}(z) \right]$ for any cyclic permutation $\{j,k,l\}$ of $\{1,2,3\}$.

We adopt the isospin-hypercharge representation~\cite{GellMann1962p1067,Jones1998} of $\mathrm{sl}(3,\mathbb{C})$ with two Cartan generators,
\begin{align}
    \bm{I}_{0} = \frac{1}{2}
    \begin{pmatrix}
        1 & 0 & 0 \\
        0 & -1 & 0 \\
        0 & 0 & 0
    \end{pmatrix}, \qquad
    \bm{Y} = \frac{1}{3}
    \begin{pmatrix}
        1 & 0 & 0 \\
        0 & 1 & 0 \\
        0 & 0 & -2
    \end{pmatrix},
\end{align}
and six ladder generators,
\begin{align}
    \begin{aligned}
    \bm{I}_{+} &= \begin{pmatrix} 0 & 1 & 0 \\ 0 & 0 & 0 \\ 0 & 0 & 0 \end{pmatrix}, &
    \bm{I}_{-} &= \begin{pmatrix} 0 & 0 & 0 \\ 1 & 0 & 0 \\ 0 & 0 & 0 \end{pmatrix}, \\
    \bm{U}_{+} &= \begin{pmatrix} 0 & 0 & 0 \\ 0 & 0 & 1 \\ 0 & 0 & 0 \end{pmatrix}, &
    \bm{U}_{-} &= \begin{pmatrix} 0 & 0 & 0 \\ 0 & 0 & 0 \\ 0 & 1 & 0 \end{pmatrix}, \\
    \bm{V}_{+} &= \begin{pmatrix} 0 & 0 & 1 \\ 0 & 0 & 0 \\ 0 & 0 & 0 \end{pmatrix}, &
    \bm{V}_{-} &= \begin{pmatrix} 0 & 0 & 0 \\ 0 & 0 & 0 \\ 1 & 0 & 0 \end{pmatrix},
    \end{aligned}
\end{align}
which correspond to the physical processes in the coupler.
The Cartan generators represent relative phase and differential gain, while the ladder generators represent coupling between mode pairs; e.g., $\bm{I}_{\pm}$ links $1 \leftrightarrow 2$, $\bm{U}_{\pm}$ links $2 \leftrightarrow 3$, and $\bm{V}_{\pm}$ links $1 \leftrightarrow 3$.
We write our effective mode-coupling matrix,
\begin{align}
    \bm{M}_{1}(z) = \sum_{X} \mu_{X}(z) \bm{X} + \sum_{X_{\pm}} \mu_{X_{\pm}}(z) \bm{X}_{\pm}, 
\end{align}
with coefficients, 
\begin{align}
    \begin{aligned}
        \mu_{X} =&~ \mathrm{Tr} \left[ \bm{M}_{1}(z) \bm{X} w_{X} \right] , \\
        \mu_{X_{\pm}} =&~ \mathrm{Tr} \left[ \bm{M}_{1}(z) \bm{X}_{\pm} \right] ,
    \end{aligned}
\end{align}
for Cartan generators $\bm{X} \in \left\{ \bm{I}_{0}, \bm{Y} \right\}$ with weights $\left\{ w_{I_{0}}, w_{Y}\right\} = \left\{2, 3/2 \right\}$, and for ladder generators $\bm{X}_{\pm} \in \left\{ \bm{I}_{\pm}, \bm{U}_{\pm}, \bm{V}_{\pm} \right\}$ with $w_{X_{\pm}} = 1$.

We extend the structure into the photonic model by promoting the Cartan and ladder generators to bosonic bilinears,
\begin{align}
    \begin{aligned}
        \hat{I}_{0} &= \frac{1}{2} ( \hat{a}_{1}^{\dagger} \hat{a}_{1} - \hat{a}_{2}^{\dagger} \hat{a}_{2} ), \\
        \hat{Y} &= \frac{1}{3} ( \hat{a}_{1}^{\dagger} \hat{a}_{1} + \hat{a}_{2}^{\dagger} \hat{a}_{2} - 2 \hat{a}_{3}^{\dagger} \hat{a}_{3} ),
    \end{aligned}
\end{align}
\begin{align}
    \begin{aligned}
        \hat{I}_{+} &= \hat{a}_{1}^{\dagger} \hat{a}_{2}, & \hat{I}_{-} &= \hat{a}_{2}^{\dagger} \hat{a}_{1}, \\
        \hat{U}_{+} &= \hat{a}_{2}^{\dagger} \hat{a}_{3}, & \hat{U}_{-} &= \hat{a}_{3}^{\dagger} \hat{a}_{2}, \\
        \hat{V}_{+} &= \hat{a}_{1}^{\dagger} \hat{a}_{3}, & \hat{V}_{-} &= \hat{a}_{3}^{\dagger} \hat{a}_{1}.
    \end{aligned}
\end{align}
Each product of ladder operators $\hat{a}_{j}^{\dagger} \hat{a}_{k}$ transfers one excitation from mode $k$ to mode $j$.

\begin{figure}[h]
    \begin{center}
        \includegraphics[scale = 1]{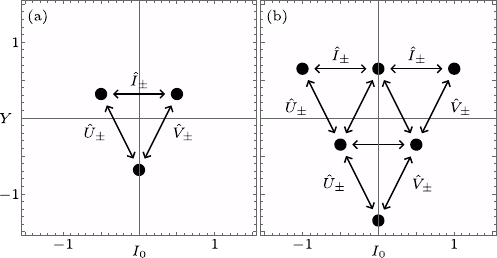}
        \caption{Weight diagrams for the totally symmetric representations $(n,0)$ in the $(I_{0},Y)$ basis, shown for (a) $n=1$ and (b) $n=2$.} \label{fig:Figure1}
    \end{center}
\end{figure}

Our effective photonic realization preserves total excitation number, allowing projection onto fixed-$n$ subspaces spanned by Fock states $\left\vert n - n_{2} - n_{3}, n_{2}, n_{3} \right\rangle$. 
Within each subspace, we adopt the isospin-hypercharge basis,
\begin{align}
    \begin{aligned}
        \hat{I}_{0} \left\vert I_{0}, Y \right\rangle &= I_{0} \left\vert I_{0}, Y \right\rangle, \\
        \hat{Y} \left\vert I_{0}, Y \right\rangle &= Y \left\vert I_{0}, Y \right\rangle,
    \end{aligned}
\end{align}
with ladder actions,
\begin{align}
    \begin{aligned}
        \hat{I}_{\pm} \left\vert I_{0}, Y \right\rangle &\propto \left\vert I_{0} \pm 1, Y \right\rangle, \\
        \hat{U}_{\pm} \left\vert I_{0}, Y \right\rangle &\propto \left\vert I_{0} \mp \frac{1}{2}, Y \pm 1 \right\rangle, \\
        \hat{V}_{\pm} \left\vert I_{0}, Y \right\rangle &\propto \left\vert I_{0} \pm \frac{1}{2}, Y \pm 1 \right\rangle.
    \end{aligned}
\end{align}
Dynkin labels \cite{Georgi2021}
\begin{align}
    (p, q) = \left( n_{1} - n_{2}, n_{2} - n_{3} \right),
\end{align}
identify irreducible representations (irreps), with the ordering $n_{1}=n-n_2-n_3 \ge n_{2} \ge n_{3}$ to ensure $p, q \ge 0$. 
We select the highest-weight state $\left\vert n, 0, 0 \right\rangle$ and work in the totally symmetric irrep $(n, 0)$ of dimension
\begin{align}
    \mathrm{dim}(n, 0) = \frac{1}{2} (n + 1)(n + 2). \label{eq:dim}
\end{align}
Each state admits two equivalent labels, given by Cartan eigenvalues or by occupation numbers,
\begin{align}
    \begin{aligned}
        \left\vert I_{0}, Y \right\rangle &= \left\vert \frac{1}{2}(n - 2 n_{2} - n_{3}), \frac{1}{3}(n - 3 n_{3}) \right\rangle \\
        &= \left\vert n - n_{2} - n_{3}, n_{2}, n_{3} \right\rangle.
    \end{aligned}
\end{align}
The states form a triangular weight diagram in the $(I_{0}, Y)$ plane with $n + 1$ layers, starting with $n + 1$ states at the top and decreasing by one per layer, Fig.~\ref{fig:Figure1}.

The single-excitation limit,  $n = 1$, reproduces the optical model as well as the differential equation system for coherent-state propagation in the non-Hermitian coupler.
The fundamental triplet,
\begin{align}
    \begin{aligned}
        \left\vert \frac{1}{2}, \frac{1}{3} \right\rangle &= \left\vert 1,0,0 \right\rangle, \\
        \left\vert -\frac{1}{2}, \frac{1}{3} \right\rangle &= \left\vert 0,1,0 \right\rangle, \\
        \left\vert 0, -\frac{2}{3} \right\rangle &= \left\vert 0,0,1 \right\rangle,
    \end{aligned}
\end{align}
spans the subspace, where optical field mode, single-photon, and coherent field amplitudes simulate quark flavor states, Fig.~\ref{fig:Figure1}(a).
For $n = 2$, the symmetric sextet,
\begin{align}
    \begin{aligned}
        \left\vert 1, \frac{2}{3} \right\rangle &= \left\vert 2,0,0 \right\rangle, &
        \left\vert \frac{1}{2}, -\frac{1}{3} \right\rangle &= \left\vert 1,0,1 \right\rangle, \\
        \left\vert 0, \frac{2}{3} \right\rangle &= \left\vert 1,1,0 \right\rangle, &
        \left\vert -\frac{1}{2}, -\frac{1}{3} \right\rangle &= \left\vert 0,1,1 \right\rangle, \\
        \left\vert -1, \frac{2}{3} \right\rangle &= \left\vert 0,2,0 \right\rangle, &
        \left\vert 0, -\frac{4}{3} \right\rangle &= \left\vert 0,0,2 \right\rangle,
    \end{aligned}
\end{align}
mirrors the structure of diquark states in hadronic physics \cite{Close1979}, Fig.~\ref{fig:Figure1}(b).
Two excitations in the three-mode coupler realize the quantum simulation of the diquark flavor multiplet, while the equivalent six-mode coupler provides its classical simulation.

The $(I_{0},Y)$ lattice defines a two-dimensional space by embedding the three-mode Fock basis into the $\mathrm{sl}(3,\mathbb{C})$ structure.  
This structure realizes synthetic dimensions~\cite{Yuan2018}, where higher-dimensional algebraic dynamics unfold within a compact three-mode system.  
The weight diagrams from this embedding show how to construct higher-dimensional classical simulations of the corresponding quantum models.

\subsection{Spectral analysis} \label{subsec:spectralanalysis}

Non-Hermitian optical and photonic systems display spectra with real, complex, or degenerate branches that determine whether propagation remains bounded, grows exponentially, or polynomially, respectively~\cite{Miri2019p7709,Meng2024p060502}. 
Spectral analysis is essential to characterize these regimes and to uncover their physical potential.

We study the spectral structure of our model within the $\mathrm{SL}(3,\mathbb{C})$ Lie group framework by diagonalizing the effective coupling matrix through similarity transformations leading to Gilmore-Perelomov coherent states \cite{Gilmore1972,Perelomov1986}.
We calculate a local spectrum, the adiabatic spectrum, by dropping the Cartan sector, and a dynamical spectrum, providing the exact spectrum, by introducing a Cartan-generated gauge that ensures coincidence with the local spectrum.

We start with the optical model and later extend the result to the photonic case~\cite{RodriguezLara2011p053845,MunozEspinosa2022p10505}, defining right and left eigenmodes
\begin{align}
    \begin{aligned}
        \bm{r}_{j}(z) =&~ \bm{T}(z) \bm{e}_{j}, \\
        \bm{l}_{j}(z) =&~ \bm{e}_{j}^{\dagger} \bm{T}^{-1}(z),
    \end{aligned}
\end{align}
for $j = 1, 2, 3$, using the basis $\left\{ \bm{e}_{1}, \bm{e}_{2}, \bm{e}_{3} \right\}$ that satisfies the biorthogonality relation
\begin{align}
    \bm{l}_{j}(z)  \bm{r}_{k}(z) = \delta_{j,k},
\end{align}
and the normal-ordered similarity transformation,
\begin{align}
    \begin{aligned}
        \bm{T}(z) &=
        e^{i \alpha_{I_{+}}(z) \bm{I}_{+}}
        e^{i \alpha_{U_{+}}(z) \bm{U}_{+}}
        e^{i \alpha_{V_{+}}(z) \bm{V}_{+}}
        e^{i \alpha_{I_{0}}(z) \bm{I}_{0}}   e^{i \alpha_{Y}(z) \bm{Y}}
        e^{i \alpha_{V_{-}}(z) \bm{V}_{-}}
        e^{i \alpha_{U_{-}}(z) \bm{U}_{-}}
        e^{i \alpha_{I_{-}}(z) \bm{I}_{-}},
    \end{aligned}
\end{align}
where $\alpha_{X}(z) \in \mathbb{C}$.
The diagonalized matrix in the similarity frame
\begin{align}
    \begin{aligned}
        \bm{M}_{D}(z) =&~ \bm{T}^{-1}(z) \bm{M}_{1}(z) \bm{T}(z) + i \bm{T}^{-1}(z) \partial_{z} \bm{T}(z) \\
        =&~ \lambda_{I_{0}}(z) \bm{I}_{0} + \lambda_{Y}(z) \bm{Y},
    \end{aligned}
\end{align}
satisfies the condition,
\begin{align}
    \bm{l}_{j}(z)  \bm{M}_{D}(z)  \bm{r}_{k}(z) = \delta_{j,k} \left[ \lambda_{I_{0}}(z) \bm{I}_{0} + \lambda_{Y}(z) \bm{Y} \right]_{j,j}.
\end{align}

At a fixed position $z = z_{0}$, the transformation $\bm{T}(z_{0})$ defines a local similarity frame
\begin{align}
    \bm{M}_{D}(z_{0}) = \bm{T}^{-1}(z_{0}) \bm{M}_{1}(z_{0}) \bm{T}(z_{0}),
\end{align}
that eliminates the dynamical term and yields a nonlinear algebraic system of eight equations,
\begin{align}
    \begin{aligned}
        \mathrm{Tr} \left[ \bm{M}_{D}(z_{0})  \bm{X} w_{X} \right] =&~ \lambda_{X}(z_{0}), \\
        \mathrm{Tr} \left[ \bm{M}_{D}(z_{0})  \bm{X}_{\pm} \right] =&~ 0,
    \end{aligned}
\end{align}
in ten variables $\left\{ \alpha_{X}(z_{0}), \alpha_{X_{\pm}}(z_{0}), \lambda_{I_{0}}(z_{0}), \lambda_{Y}(z_{0}) \right\}$.
We close the system by fixing the local gauge $\alpha_{I_{0}} = \alpha_{Y} = 0$, which yields the local spectrum,
\begin{align}
    \begin{aligned}
        \lambda_{I_{0}}(z_{0}) =&~ \lambda_{1}(z_{0}) - \lambda_{2}(z_{0}), \\
        \lambda_{Y}(z_{0}) =&~ \frac{3}{2} \left[ \lambda_{1}(z_{0}) + \lambda_{2}(z_{0}) \right] = - \frac{3}{2} \lambda_{3}(z_{0}),
    \end{aligned}
\end{align}
in terms of the roots $\{ \lambda_{1}(z_{0}), \lambda_{2}(z_{0}), \lambda_{3}(z_{0}) 
= -\lambda_{1}(z_{0}) - \lambda_{2}(z_{0}) \}$ of the depressed cubic
\begin{align}
    \lambda_{i}^{3}(z_{0}) - \frac{1}{2} \mathrm{Tr} \left[ \bm{M}_{1}^{2}(z_{0}) \right] \lambda_{i}(z_{0}) - \frac{1}{3} \mathrm{Tr} \left[ \bm{M}_{1}^{3}(z_{0}) \right] = 0. \label{eq:CharPol}
\end{align}

For the dynamical case, we enforce the spectrum to match the local eigenvalues at all positions,
\begin{align}
    \begin{aligned}
        \lambda_{I_{0}}(z) =&~ \lambda_{1}(z) - \lambda_{2}(z), \\
        \lambda_{Y}(z) =&~  \frac{3}{2} \left[ \lambda_{1}(z) + \lambda_{2}(z) \right] = - \frac{3}{2} \lambda_{3}(z),
    \end{aligned}
\end{align}
with the roots satisfying the corresponding dynamical depressed cubic.
The dynamical diagonalization equation, 
\begin{align}
    & \bm{T}^{-1}(z) \bm{M}_{1}(z) \bm{T}(z) + i \bm{T}^{-1}(z) \partial_{z} \bm{T}(z) \\
    & = \lambda_{I_{0}}(z) \bm{I}_{0} + \lambda_{Y}(z) \bm{Y}, \nonumber
\end{align}
yields eight coupled differential equations, 
\begin{align}
    \begin{aligned}
        \mathrm{Tr} \left[ \bm{M}_{D}(z)  \bm{X} w_{X} \right] =&~ \lambda_{X}(z), \\
        \mathrm{Tr} \left[ \bm{M}_{D}(z)  \bm{X}_{\pm} \right] =&~ 0,
    \end{aligned}
\end{align}
in eight variables $\left\{ \alpha_{X}(z), \alpha_{X_{\pm}}(z) \right\}$.

The Cartan parameters $\left\{ \alpha_{I_{0}}(z), \alpha_{Y}(z) \right\}$ define a dynamical gauge that absorbs the contribution $i \bm{T}^{-1}(z) \partial_{z} \bm{T}(z)$.
Their propagation encodes a geometric phase~\cite{Berry1984p45} determined by the path traced in parameter space and originates from the topology of the underlying symmetry~\cite{Wilczek1984p2111}.
This phase is the holonomy~\cite{Simon1983p211}, 
\begin{align}
\bm{H} =  \mathrm{Pexp} \left[ i \oint_{\mathcal{C}} \mathrm{d}z  \, \bm{T}^{-1}(z) \partial_{z} \bm{T}(z) \right],
\end{align}
for a closed path $\mathcal{C}$, where $\mathrm{Pexp}$ is a path-ordered exponential.
Even if the local frame is cyclic, the dynamical frame may undergo a nontrivial transformation governed by $\bm{H}$, revealing a gauge-dependent geometric phase associated with the Cartan sector \cite{MehriDehnavi2008p082105}. 
Geometric phases unify holonomy, topology, and environment-induced dynamics, and play a central role in condensed matter physics, topological transitions in open quantum platforms \cite{FerrerGarcia2023p6810}, 
and prospective quantum technologies \cite{Xu2012p170501}.

We extend the optical result to the photonic model by promoting the similarity transformation and gauge connection to operators in the bilinear bosonic representation.
The diagonal mode-coupling operator,
\begin{align}
    \begin{aligned}
        \hat{M}_{D}(z) =&~ \lambda_{I_{0}}(z) \hat{I}_{0} + \lambda_{Y}(z) \hat{Y} \\
                       =&~ \frac{1}{2} \left[ 3 \lambda_{1}(z) - \lambda_{2}(z) \right] \hat{n} + 2 \left[ \lambda_{2}(z)  - \lambda_{1}(z) \right] \hat{n}_{2}  - \frac{1}{2} \left[ 5 \lambda_{1}(z) + \lambda_{2}(z) \right] \hat{n}_{3},
    \end{aligned}
\end{align}
acts in terms of the total and modal excitation numbers.
The right and left eigenstates,
\begin{align}
    \begin{aligned}
        \left\vert r_{n, n_{2}, n_{3}}(z) \right\rangle =&~ \hat{T}(\zeta) \left\vert n - n_{2} - n_{3}, n_{2}, n_{3} \right\rangle, \\
        \left\langle l_{n, n_{2}, n_{3}}(z) \right\vert =&~ \left\langle n - n_{2} - n_{3}, n_{2}, n_{3} \right\vert \hat{T}^{-1}(\zeta),
    \end{aligned}
\end{align}
follow from the similarity transformation and its inverse, where $\zeta = z_{0}$ for the local and $\zeta = z $ for the dynamical frame.

In practice, the similarity transformation and its inverse are group elements in $\mathbf{SL}(3, \mathbb{C})$ that  can be decomposed back into the isospin-hypercharge representation,
\begin{align}
    \begin{aligned}
        \bm{T}(\zeta)      =&~ \tau_{1} \bm{1} + \sum_{X} \tau_{X}(\zeta) \bm{X}, \\
        \bm{T}^{-1}(\zeta) =&~ \sigma_{1} \bm{1} + \sum_{X} \sigma_{X}(\zeta) \bm{X},
    \end{aligned}
\end{align}
with $X \in \left\{ \bm{I}_{0}, \bm{Y}, \bm{I}_{\pm}, \bm{U}_{\pm}, \bm{V}_{\pm} \right\}$ and promote each generator $\bm{X}$ to its bilinear bosonic operator $\hat{X}$,
\begin{align}
    \begin{aligned}
        \left\vert r_{n, n_{2}, n_{3}}(z) \right\rangle =&~ \sum_{X} \, \tau_{X}(\zeta) \hat{X} \left\vert n - n_{2} - n_{3}, n_{2}, n_{3} \right\rangle, \\
        \left\langle l_{n, n_{2}, n_{3}}(z) \right\vert =&~ \sum_{X} \sigma_{X}(\zeta)  \, \left\langle n - n_{2} - n_{3}, n_{2}, n_{3} \right\vert \hat{X},
    \end{aligned}
\end{align}
for analytical insight and numerical implementation.

\subsection{Propagation dynamics} \label{subsec:propdyn}

While
spectral analysis provides useful information to characterize the system, understanding the full dynamics requires an explicit propagator.
We construct an explicit propagator using a normal-ordered Wei–Norman decomposition \cite{Wei1963p575} in terms of the $\mathrm{SL}(3,\mathbb{C})$ elements,
\begin{align}\label{propagation}
    \begin{aligned}
        \bm{U}(z) &= e^{i \upsilon_{I_{+}}(z) \bm{I}_{+}}
                    e^{i \upsilon_{U_{+}}(z) \bm{U}_{+}}
                    e^{i \upsilon_{V_{+}}(z) \bm{V}_{+}}
                    e^{i \upsilon_{I_{0}}(z) \bm{I}_{0}}   e^{i \upsilon_{Y}(z) \bm{Y}}
                    e^{i \upsilon_{V_{-}}(z) \bm{V}_{-}}
                    e^{i \upsilon_{U_{-}}(z) \bm{U}_{-}}
                    e^{i \upsilon_{I_{-}}(z) \bm{I}_{-}},
    \end{aligned}
\end{align}
which satisfies
\begin{align}
    \partial_{z} \bm{U}(z) = i \bm{M}_{1}(z) \bm{U}(z), \qquad \bm{U}(0) = \bm{1},
\end{align}
such that $\upsilon_{X}(0) = 0$ for all generators.
The propagation of an initial state,
\begin{align}
    \begin{aligned}
        \bm{E}(z) =&~ \bm{U}(z) \bm{E}(0) \\
                  =&~ \sum_{j=1}^{3} E_{j}(0) \, \bm{U}(z) \bm{e}_{j},
    \end{aligned}
\end{align}
is a coherent superposition of Gilmore–Perelomov states generated by the group action.

Substituting the decomposition into the propagation equation yields a triangular system~\cite{Charzynski2013p265208} composed of a coupled Riccati pair,
\begin{align}
    \begin{pmatrix}
        \upsilon_{I_{+}}^{\prime} \\
        \upsilon_{V_{+}}^{\prime}
    \end{pmatrix}
    =&~
    \begin{pmatrix}
        \mu_{12} \\
        \mu_{13}
    \end{pmatrix}
    + i
    \begin{pmatrix}
        \mu_{11} - \mu_{22} & -\mu_{32} \\
        -\mu_{23} & 2 \mu_{11} + \mu_{22}
    \end{pmatrix}
    \begin{pmatrix}
        \upsilon_{I_{+}} \\
        \upsilon_{V_{+}}
    \end{pmatrix} \nonumber \\
    &~
    +
    \begin{pmatrix}
        \upsilon_{I_{+}} \\
        \upsilon_{V_{+}}
    \end{pmatrix}
    \begin{pmatrix}
        \mu_{21} & \mu_{31}
    \end{pmatrix}
    \begin{pmatrix}
        \upsilon_{I_{+}} \\
        \upsilon_{V_{+}}
    \end{pmatrix},
\end{align}
a scalar Riccati equation,
\begin{align}
    \begin{aligned}
        \upsilon_{U_{+}}^{\prime}
        =&~ \mu_{23} + i \mu_{21} \upsilon_{V_{+}} + \left( i \mu_{11} + 2i \mu_{22} - \mu_{21} \upsilon_{I_{+}} + \right. \\
        &~ \left. + \mu_{31} \upsilon_{V_{+}} \right) \upsilon_{U_{+}}  + \left( \mu_{32} + i \mu_{31} \upsilon_{I_{+}} \right) \upsilon_{U_{+}}^{2},
    \end{aligned}
\end{align}
and linear equations for the remaining variables,
\begin{align}
    \begin{aligned}
        \upsilon_{I_{0}}^{\prime} =&~ \mu_{11} - \mu_{22} - \mu_{31} \upsilon_{I_{+}} \upsilon_{U_{+}} \\
                                  &~ - i \left( 2 \mu_{21} \upsilon_{I_{+}} + \mu_{31} \upsilon_{V_{+}} - \mu_{32} \upsilon_{U_{+}} \right), \\
        \upsilon_{Y}^{\prime} =&~ \frac{3}{2} \left[ \mu_{11} + \mu_{22} - i \left( \mu_{32} \upsilon_{U_{+}} + \mu_{31} \upsilon_{V_{+}} \right) + \mu_{31} \upsilon_{I_{+}} \upsilon_{U_{+}} \right], \\
        \upsilon_{V_{-}}^{\prime} =&~ e^{\frac{i}{2} \upsilon_{I_{0}}} \left[ \mu_{31} e^{i \upsilon_{Y}} - e^{\frac{i}{2} \upsilon_{I_{0}}} \left( i \mu_{21} + \mu_{31} \upsilon_{U_{+}} \right) \upsilon_{U_{-}} \right], \\
        \upsilon_{U_{-}}^{\prime} =&~ e^{ -\frac{i}{2} \upsilon_{I_{0}} + i \upsilon_{Y} } \left( \mu_{32} + i \mu_{31} \upsilon_{I_{+}} \right), \\
        \upsilon_{I_{-}}^{\prime} =&~ e^{ i \upsilon_{I_{0}} } \left( \mu_{21} - i \mu_{31} \upsilon_{U_{+}} \right).
    \end{aligned}
\end{align}
The hierarchy admits sequential integration: solve the Riccati pair $\left\{ \upsilon_{I_{+}}, \upsilon_{V_{+}} \right\}$, substitute into the scalar equation for $\upsilon_{U_{+}}$, then integrate the remaining variables. 
Local existence and uniqueness follow from analyticity provided $\mu_{ij}(z)$ are continuous over the integration interval~\cite{Wei1964p327}. 

We promote the classical propagator, Eq.~\eqref{propagation}, to a photonic operator,
\begin{align}
    \hat{U}(z) =&~ e^{i \upsilon_{I_{+}}(z) \hat{I}_{+}}
                 e^{i \upsilon_{U_{+}}(z) \hat{U}_{+}}
                 e^{i \upsilon_{V_{+}}(z) \hat{V}_{+}}
                 e^{i \upsilon_{I_{0}}(z) \hat{I}_{0}}
                 e^{i \upsilon_{Y}(z) \hat{Y}}
                 e^{i \upsilon_{V_{-}}(z) \hat{V}_{-}}
                 e^{i \upsilon_{U_{-}}(z) \hat{U}_{-}}
                 e^{i \upsilon_{I_{-}}(z) \hat{I}_{-}} .
\end{align}
driven by the classical envelopes $\upsilon_{X}(z)$ that solve the Wei-Norman system.
It satisfies 
\begin{align}
    \partial_{z} \hat{U}(z) = i \hat{M}_{1}(z) \hat{U}(z), \qquad \hat{U}(0) = \hat{1},
\end{align}
and propagates states, 
\begin{align}
    \left\vert E(z) \right\rangle = \hat{U}(z) \left\vert E(0) \right\rangle,
\end{align}
with initial conditions,
\begin{align}
    \vert E(0) \rangle = \sum_{n_{2},n_{3}} c_{n, n_{2},n_{3}} \left\vert n - n_{2} - n_{3} , n_{2}, n_{3} \right\rangle,
\end{align}
restricted to fixed-$n$ subspaces.

In photonic systems, we track the mode excitation numbers, 
\begin{align}
    \begin{aligned}
        \hat{n}_{1}(z) =&~ \frac{1}{3} \hat{n} + \frac{1}{2} \hat{Y}(z) + \hat{I}_{0}(z), \\
        \hat{n}_{2}(z) =&~ \frac{1}{3} \hat{n} + \frac{1}{2} \hat{Y}(z) - \hat{I}_{0}(z), \\
        \hat{n}_{3}(z) =&~ \frac{1}{3} \hat{n} - \hat{Y}(z),
    \end{aligned}
\end{align}
which reduce to the Cartan sector, 
\begin{align}
    \begin{aligned}
        \hat{I}_{0}(z) =&~\hat{U}^{-1}(z) \hat{I}_{0} \hat{U}(z) &=&~\sum_{X} \iota_{X}(z) \hat{X}, \\
        \hat{Y}(z)     =&~\hat{U}^{-1}(z) \hat{Y}     \hat{U}(z) &=&~ \sum_{X} \eta_{X}(z) \hat{X}, 
    \end{aligned}
\end{align}
since the total excitation number $\hat{n}$ is a constant of propagation.
The closed form for the coefficients $\iota_{X}(z)$ and $\eta_{X}(z)$ 
as functions of the classical envelopes $\upsilon_{X}(z)$ follow from the 
Baker-Campbell-Hausdorff identities; we omit them for brevity.

The propagator traces a trajectory on the $\mathrm{SL}(3,\mathbb{C})$ group manifold with the Wei–Norman parameters $\upsilon_{X}(z)$ as coordinates along Cartan and ladder directions.
The spectrum fixes the local tangent, while integration generates the orbit that encodes the full dynamics.
In this way, propagation dynamics complements the spectral analysis by turning eigenvalue structure into explicit state evolution.

\subsection{Higher-order exceptional points} \label{subsec:hoeps}

According to Eq.~\eqref{eq:CharPol},
the eigenvalues of the effective $\mathrm{sl}(3,\mathbb{C})$ 
coupler satisfy the depressed cubic,
\begin{align}
    \lambda^{3}(z)+\beta_{2}(z)\lambda(z)+\beta_{3}(z)=0
\end{align}
with coefficients given by the quadratic ($j=2$) and cubic ($j=3$) invariants,
\begin{align}
        \beta_{j}(z) =~ -\frac{1}{j} \mathrm{Tr}\, \left[ \bm{M}_{1}^{j}(z) \right],    
\end{align}
of the traceless mode-coupling matrix $\bm{M}_{1}(z)$. 
The discriminant,
\begin{align}
    \Delta(z)=-4 \beta_{2}^{3}(z) - 27 \beta_{3}^{2}(z),
\end{align}
classifies the spectral regimes. 
If $\Delta(z)\neq0$, all eigenvalues are distinct. 
If $\Delta(z)=0$ with $\{\beta_{2}(z),\beta_{3}(z)\}\neq\{0,0\}$, two eigenvalues collapse, giving a second-order exceptional point (EP2). 
If $\beta_{2}(z)=\beta_{3}(z)=0$, the matrix is nilpotent of index three, all eigenvalues collapse, and a third-order exceptional point (EP3) arises. 
These conditions parcel parameter space into the discriminant surface for EP2 and the nilpotent cone for EP3  \cite{Fulton2004}, connecting the matrix invariants and the topology of the spectral Riemann surface.

The photonic spectral structure follows from the optical one, and the discriminant defines identical regions in both cases.
In the optical model, an EP$m$ arises when the mode-coupling matrix has a Jordan canonical form with block size $m$.
Embedding an $n$-excitation photonic sector into the $\mathrm{sl}(3,\mathbb{C})$ structure produces a Fock subspace with dimension $(n+1)(n+2)/2$, Eq.~\ref{eq:dim}.
The single Jordan block of size $m$ in the optical model becomes an operator with nilpotency of index $(m-1)n + 1$ in the Fock subspace. 
Thus, the parameter values that yield EP2 and EP3 in the optical coupler correspond to EP$(n+1)$ and EP$(2n+1)$ in the photonic case, and we recover the optical limit for $n=1$ as expected.

\begin{figure}[t]
    \begin{center}
 \includegraphics[scale = 0.7]{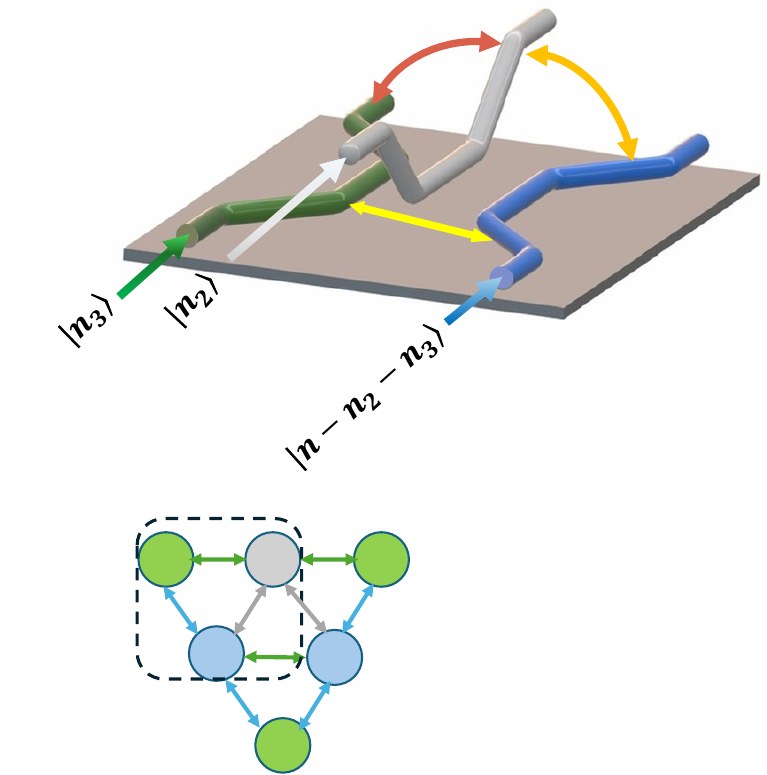}  
 \caption{Schematic of a single, lossy waveguide beam splitter with three waveguides excited with $N$ indistinguishable photons prepared in the state $\ket{ n - n_{2} - n_{3}, n_{2}, n_{3}}$. In the Schwinger representation, it can be mapped to a kagome lattice \cite{Sankar2024p021046}. The coupling between adjacent “modes” is given by matrix elements of $\bm{I}_{\pm}$, $\bm{U}_{\pm}$ and $\bm{V}_{\pm}$.}   \label{fig:Figuresetup}  
    \end{center}
\end{figure}

\section{\texorpdfstring{$\mathcal{PT}$}{PT}-symmetric and non-Hermitian cyclic couplers} \label{sec:PTsymcc}

\begin{figure}[h]
    \begin{center}
        \includegraphics[scale = 1]{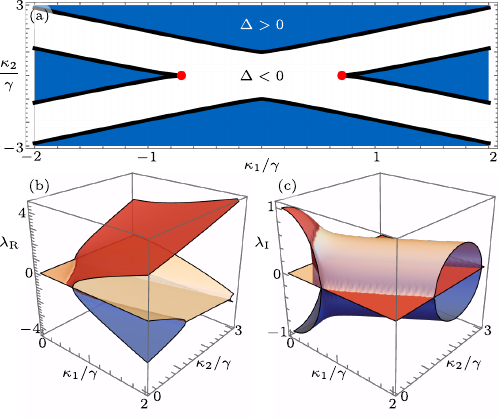}
        \caption{Spectral structure of the non-Hermitian coupler family in Eq.~(\ref{eq:Example}).
        (a) Sign of the discriminant with $\Delta>0$ in blue, $\Delta<0$ in white, $\Delta=0$ in black for EP2, and red for EP3. 
        (b) Real and (c) imaginary part of the spectral Riemann surface with branches in red, cream, and blue.}
        \label{fig:Figure2}  
    \end{center}
\end{figure}

As an example of our general framework developed in Sec.~\ref{sec:Sec2},
we study the non-Hermitian coupler family depicted schematically in 
Fig.~\ref{fig:Figuresetup},
\begin{align}
    \begin{aligned}
        \bm{M}_{1}(z) =&~
            \begin{pmatrix}
                i \gamma(z) & \kappa_{1}(z) & \kappa_{2}(z) \\
                \kappa_{1}(z)   & 0         & \kappa_{1}(z) \\
                \kappa_{2}(z)   & \kappa_{1}(z) & -i \gamma(z)
            \end{pmatrix} \\
        =&~ i \gamma(z) \left( \bm{I}_{0} + \tfrac{3}{2} \bm{Y} \right)  + \kappa_{1}(z) \left( \bm{I}_{+} + \bm{I}_{-} + \bm{U}_{+} + \bm{U}_{-} \right) +  \kappa_{2}(z) \left( \bm{V}_{+} + \bm{V}_{-} \right),
    \end{aligned} \label{eq:Example}
\end{align}
with $\gamma, \kappa_{j} \in \mathbb{R}$.
We do not introduce three independent couplings, as EP3s only exist when the ladder sectors $\bm{I}_{\pm}$ and $\bm{U}_{\pm}$ share the same coupling.  
$\bm{M}_{1}(z)$ in Eq.~\eqref{eq:Example}  
reduces to the $\mathcal{PT}$-symmetric trimer \cite{ZaragozaGutierrez2016p3989} when $\kappa_{2}(z)=0$, with EP3 at $\gamma^{2}=2\kappa_{1}^{2}$, and to a cyclic trimer when $\kappa_{1}(z)=\kappa_{2}(z)=\kappa(z)$, without exceptional points.  
The invariants,
\begin{align}
    \begin{aligned}
        \beta_{2}(z) =&~ \gamma^{2}(z) - 2 \kappa_{1}^{2}(z) - \kappa_{2}^{2}(z), \\
        \beta_{3}(z) =&~ - 2 \kappa_{1}^{2}(z) \kappa_{2}(z),
    \end{aligned}
\end{align}
lead to the discriminant,
\begin{align}
    \begin{aligned}
        \Delta(z) =&~ -4 \left[ \gamma^{2}(z) - 2 \kappa_{1}^{2}(z) - \kappa_{2}^{2}(z) \right]^{3} - 27 \left[ 2 \kappa_{1}^{2}(z) \kappa_{2}(z) \right]^{2},        
    \end{aligned}
\end{align}
which defines the spectral structure, Fig.~\ref{fig:Figure2}.
If $\Delta(z)\neq0$, the three eigenvalues are distinct and no exceptional point arises.  
If $\Delta(z)=0$ with $\{\beta_{2}(z),\beta_{3}(z)\}\neq\{0,0\}$, two eigenvalues and their eigenvectors coalesce, yielding EP2s shown as black dots in Fig.~\ref{fig:Figure2}(a).  
At $\beta_{2}(z)=\beta_{3}(z)=0$, the matrix is nilpotent, all eigenvalues collapse to zero, and two EP3s emerge, shown as red dots in Fig.~\ref{fig:Figure2}(a).  
Thus, this non-Hermitian coupler family admits regimes without degeneracy, surfaces of EP2s, and EP3s on the nilpotent cone, excluding the trivial zero matrix.
In the $\mathcal{PT}$-symmetric case, the spectrum transitions from real to complex-conjugate pairs across an EP3, while in the cyclic case the only degeneracy is the diabolical point at $\gamma(z)=0$, where the spectrum shows a Hermitian conical intersection \cite{Berry1984p15}.  

Inside the EP2 curves, white region in Fig.~\ref{fig:Figure2}(a), the discriminant is negative and the spectrum has one real eigenvalue and a complex-conjugate pair,  Figs.~\ref{fig:Figure2}(b) and \ref{fig:Figure2}(c),  producing non-compact orbits with
hyperbolic dynamics. 
Outside the curves, blue region in Fig.~\ref{fig:Figure2}(a), the discriminant is positive and all three eigenvalues are real and distinct, producing compact orbits with periodic dynamics.

\begin{figure}[h]
    \begin{center}
        \includegraphics[scale = 1]{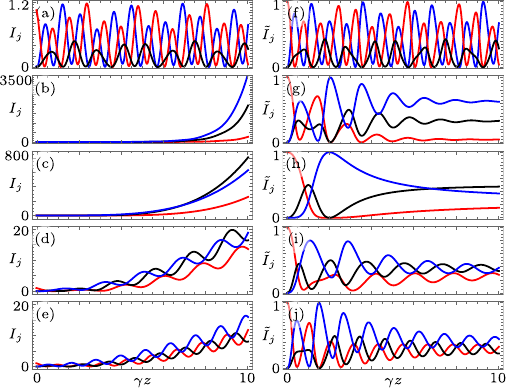}
        \caption{Dynamics of the non-Hermitian coupler family in Eq.~\ref{eq:Example} for the initial classical field $\bm{E}(0) = \left\{1, 0, 0 \right\}$. 
        (a)–(e) Classical field intensities and (f)–(j) their renormalization, showing   
        (a,f) compact periodic dynamics for $\Delta>0$ with $\kappa_{1} / \gamma = 3/2$, $\kappa_{2} / \gamma = 7/2$;  
        (b,g) non-compact hyperbolic dynamics for $\Delta<0$ with $\kappa_{1} / \gamma = \kappa_{2} / \gamma = 3/2$;  
        (c,h) quadratic algebraic dynamics at EP3 with $\Delta=0$, $\kappa_{1} / \gamma = 1 / \sqrt{2}$, $\kappa_{2} / \gamma = 0$;  
        (d,i) and (e,j) linear dynamics at EP2 with $\Delta=0$, $\kappa_{1} / \gamma = 3/2$, $\kappa_{2} / \gamma \in \{0.6718,2.3882\}$.
        Solid lines in
        red, black, and blue correspond to the intensities of the field amplitudes $E_{1}(z)$, $E_{2}(z)$, and $E_{3}(z)$, respectively.}
        \label{fig:Figure3}  
    \end{center}
\end{figure}

The dynamics in Fig.~\ref{fig:Figure3} use the initial condition $\bm{E}(0)=\{1,0,0\}$, which corresponds to a classical field, a single excitation state, or a coherent state input impinging at the first element.  
We show intensities
\begin{align}
    I_{j}(z) = \vert E_{j}(z) \vert^{2}
\end{align}
in Fig.~\ref{fig:Figure3}(a)--(e) and their renormalization 
\begin{align}
    \tilde{I}_{j}(z) = \frac{ \vert E_{j}(z) \vert^{2} } { \sum_{j=1}^{3} \vert E_{j}(z) \vert^{2} }    
\end{align}
in Fig.~\ref{fig:Figure3}(f)--(j).
The eigenvalues determine whether the propagator contains oscillatory, polynomial, or hyperbolic terms, which translate into compact or non-compact orbits on the group manifold.
For $\Delta>0$, all eigenvalues are real, the propagator is a superposition of oscillatory terms, and the dynamics follows compact orbits with bounded oscillations, Fig.~\ref{fig:Figure3}(a) and Fig.~\ref{fig:Figure3}(f).  
For $\Delta<0$, one eigenvalue is real and the other two are a complex-conjugate pair, the propagator mixes oscillatory and hyperbolic terms, and the dynamics follows non-compact orbits with unbounded hyperbolic growth, Fig.~\ref{fig:Figure3}(b) and Fig.~\ref{fig:Figure3}(g).  
At EP3 all eigenvalues collapse to zero, the matrix is nilpotent of index three, thus the propagator is quadratic in $z$ \cite{ZaragozaGutierrez2016p3989}, and the dynamics follows non-compact orbits with unbounded quadratic growth, Fig.~\ref{fig:Figure3}(c) and Fig.~\ref{fig:Figure3}(h).
At EP2 two eigenvalues and eigenvectors coalesce, the propagator mixes linear, oscillatory, and hyperbolic terms, and the dynamics follow non-compact orbits with unbounded linear, Fig.~\ref{fig:Figure3}(d) and Fig.~\ref{fig:Figure3}(i), and hyperbolic growth, Fig.~\ref{fig:Figure3}(e) and Fig.~\ref{fig:Figure3}(j).
All cases except the fourth, Fig.~\ref{fig:Figure3}(d) and Fig.~\ref{fig:Figure3}(i), show transfer between the first and third renormalized modes.  

Previous studies of non-Hermitian optical and photonic systems analyzed propagation-dependent trajectories on the spectral Riemann surface that either encircle or cross exceptional points.  
Encircling an EP of order $m$ produces a cyclic permutation of eigenvalue branches, a monodromy, with full return to the initial branch after $m$ loops \cite{Holler2020p032216,Sayyad2022p023130}.  
Crossing an EP instead collapses dynamics, with the post-crossing state depending on the approach path, rate and symmetry-breaking perturbations \cite{Zhong2019p134}.  
For our non-Hermitian family, pure encircling is ruled out; any loop around an EP3 in the $(\kappa_{1},\kappa_{2})$ plane also encloses infinitely many EP2s and necessarily crosses two of them.  

To explore propagation-dependent dynamics, we consider circular loops in the $(\kappa_{1},\kappa_{2})$ plane,
\begin{align}
    \begin{aligned}
        \frac{\kappa_{1}(z)}{\gamma(z)} &= \frac{1}{\sqrt{2}} + r \cos\!\left[ 2 \pi \gamma(z) z \right], \\
        \frac{\kappa_{2}(z)}{\gamma(z)} &= r \sin\!\left[ 2 \pi \gamma(z) z \right],    
    \end{aligned}
\end{align} 
with $\gamma(z) z \in [0,m]$ parametrizing $m$ clockwise turns of radius $r$ around EP3 at $( \kappa_{1}(z) / \gamma(z), \kappa_{2}(z) / \gamma(z) ) = (1 / \sqrt{2}, 0 )$.  

\begin{figure}[h]
    \begin{center}
        \includegraphics[scale = 1]{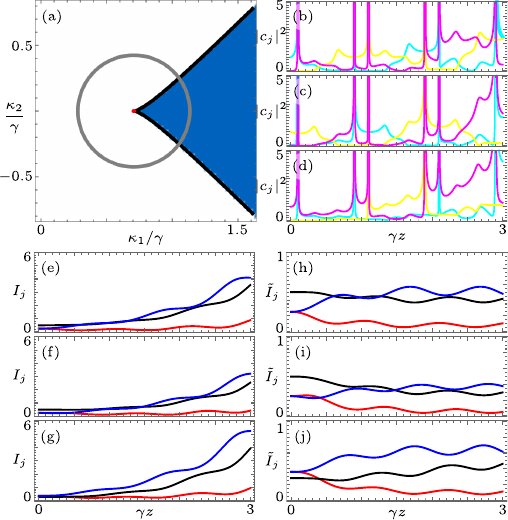}
        \caption{Propagation-dependent dynamics for three clockwise turns along a circular loop in the $(\kappa_{1},\kappa_{2})$ plane centered at EP3.  
        (a) Sign of the discriminant with the trajectory.  
        (b)–(d) Projection into the instantaneous right eigenbasis with $|c_{j}(z)|^{2}$ in cyan, yellow, and magenta for the first, second, and third eigenvector.  
        (e)–(g) Field intensities and (h)–(j) their renormalization in red, black, and blue, as in Fig.~\ref{fig:Figure3}.} 
        \label{fig:Figure4}
    \end{center}
\end{figure}

Figure~\ref{fig:Figure4} shows propagation-dependent dynamics for three clockwise turns of radius $r=0.4253$ around EP3, Fig.~\ref{fig:Figure4}(a). 
We decompose the field into the instantaneous biorthogonal basis,
\begin{align}
    c_{j}(z) = \bm{l}_{j}(z)\bm{E}(z),
\end{align}
and show the coefficients $|c_{j}(z)|^{2}$ in Fig.~\ref{fig:Figure4}(b)--Fig.~\ref{fig:Figure4}(d) for initial states given by the three right eigenmodes at $z=0$. Since any loop enclosing an EP3 also encloses infinitely many EP2 and necessarily crosses two of them, pure encircling is not possible and no monodromy arises.  
Each EP2 crossing collapses two eigenmodes, producing sharp peaks in their coefficients while the third remains smooth.
Subsequent propagation mixes the basis and all components exhibit signatures of the following crossings.  
The intensities, Fig.~\ref{fig:Figure4}(e)--Fig.~\ref{fig:Figure4}(g), show how crossings alter the dynamics of the physical modes. 
In the compact regime $\Delta > 0$, oscillations are bounded, but along the circular loop the fields enter non-compact regimes $\Delta < 0$ and undergo amplification or decay. 
Renormalization, Fig.~\ref{fig:Figure4}(h)--Fig.~\ref{fig:Figure4}(j), removes the overall amplification or decay and exposes the relative exchange between the modes.

\begin{figure}[h]
    \begin{center}
        \includegraphics[scale = 1]{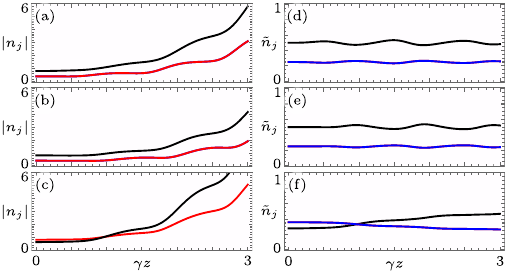}
        \caption{Propagation-dependent dynamics for three clockwise turns along a circular loop in the $(\kappa_{1},\kappa_{2})$ plane centered at EP3.  
        (a)–(c) Biorthogonal mode populations $|n_j(z)|$ for three initial right eigenmodes at $z=0$ and their renormalization (d)–(f) $\tilde{n}_{j}$.} 
        \label{fig:Figure5}
    \end{center}
\end{figure}

A true non-Hermitian quantum system would require calculating observables with biorthogonal expectation values; e.g., the mode excitation numbers, 
\begin{align}
    n_{j} = \left\langle l_{n_{1}, n_{2}, n_{3} }(z) \right\vert \hat{n}_{j} \left\vert r_{n_{1}, n_{2}, n_{3}}(z)   \right\rangle \in \mathbb{C},
\end{align} 
and its renormalization,
\begin{align}
    \tilde{n}_{j} = \frac{ n_{j} }{\sum_{j=1}^{3} \vert n_{j} \vert  } \in \mathbb{C},
\end{align} 
that even in the single-excitation limit does not coincide with the classical result, Fig.~\ref{fig:Figure5}.
Fundamental closed-system Hamiltonians are Hermitian; non-Hermitian dynamics arise effectively via loss/gain engineering, measurements, and post-selection. 

Lossy photonic dynamics reproduces effective non-Hermitian evolution only in the single-excitation limit, where Lindblad dynamics coincides with the non-Hermitian description \cite{JaramilloAvila2020p1761}. 
For higher excitation numbers, the open-system dynamics departs from the non-Hermitian picture.
Indeed, in open quantum systems, the Lindblad and Langevin formalisms provide rigorous reductions of unitary dynamics generated by a Hermitian Hamiltonian. 
The Lindblad equation preserves complete positivity, while the Langevin approach yields an effective non-Hermitian Hamiltonian supplemented by stochastic noise operators. 
The commonly used non-Hermitian Schr\"odinger equation corresponds to the noise-averaged limit of the Langevin formalism, valid only at the mean-field or single-excitation level. 
Genuine realization of such effective non-Hermitian dynamics requires post-selection within an open-system framework.
Access to the effective exceptional-point structure requires excitation-resolved detection for post-selection in the corresponding Fock subspace \cite{QuirozJuarez2019p862}.
In these cases, we can reconstruct the amplitudes, 
\begin{align}
    P_{n_{1},n_{2},n_{3}}(z) = \vert \left\langle n_{1}, n_{2}, n_{3} \right\vert \left. E(z)   \right\rangle \vert^{2},
\end{align}
and their renormalization, 
\begin{align}
    \tilde{P}_{n_{1},n_{2},n_{3}}(z) = \frac{P_{n_{1},n_{2},n_{3}}(z)}{\sum_{n_{1},n_{2},n_{3}} P_{n_{1},n_{2},n_{3}}(z)} ,
\end{align}
which in the single-excitation limit coincide with the classical result, Fig.~\ref{fig:Figure4}(e)--Fig.~\ref{fig:Figure4}(j).

\begin{figure}[h]
    \begin{center}
        \includegraphics[scale = 1]{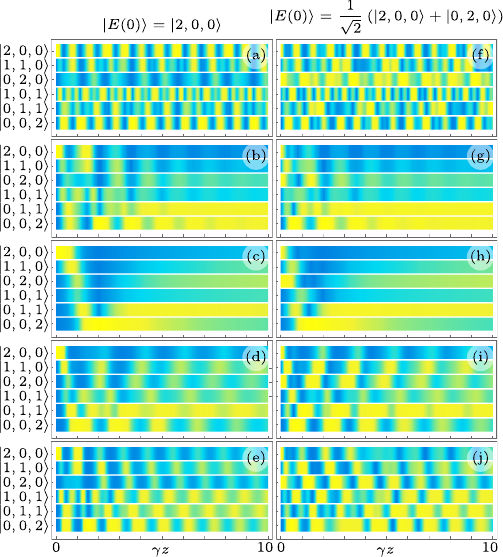}
        \caption{Renormalized amplitudes $\tilde{P}_{n_{1},n_{2},n_{3}}$ of our non-Hermitian coupler family in Eq.~\ref{eq:Example} for the initial photonic states (a)–(e) $\vert {E}(0) \rangle = \left\vert 2, 0, 0 \right\rangle$ and (f)–(j) the NOON state $\vert {E}(0) \rangle = \left( \left\vert 2, 0, 0  \right\rangle + \left\vert 0, 2, 0  \right\rangle \right)/\sqrt{2}$.
        (a,f) compact periodic dynamics for $\Delta>0$ with $\kappa_{1} / \gamma = 3/2$, $\kappa_{2} / \gamma = 7/2$;  
        (b,g) non-compact hyperbolic dynamics for $\Delta<0$ with $\kappa_{1} / \gamma = \kappa_{2} / \gamma = 3/2$;  
        (c,h) quadratic algebraic dynamics at EP3 with $\Delta=0$, $\kappa_{1} / \gamma = 1 / \sqrt{2}$, $\kappa_{2} / \gamma = 0$;  
        (d,i) and (e,j) linear dynamics at EP2 with $\Delta=0$, $\kappa_{1} / \gamma = 3/2$, $\kappa_{2} / \gamma \in \{0.6718,2.3882\}$.} 
        \label{fig:Figure7}
    \end{center}
\end{figure}
Figure~\ref{fig:Figure7} presents the renormalized amplitudes \( P_{n_{1}, n_{2}, n_{3}}(z) \) for two-excitation inputs in our non-Hermitian coupler family defined in Eq.~\ref{eq:Example}. 
We examine the propagation dynamics for two types of initial states: a Fock state \( \vert 2,0,0 \rangle \), Figs.~\ref{fig:Figure7}(a)--Figs.~\ref{fig:Figure7}(e), and a NOON state \( \left( \vert 2,0,0 \rangle + \vert 0,2,0 \rangle \right)/\sqrt{2} \), Figs.~\ref{fig:Figure7}(f)--Figs.~\ref{fig:Figure7}(j). 
The results illustrate the system behavior across the four spectral regimes defined in Fig.~\ref{fig:Figure3}: compact periodic oscillations for \( \Delta > 0 \), Fig.~\ref{fig:Figure7}(a) and Figs.~\ref{fig:Figure7}(f); non-compact hyperbolic growth for \( \Delta < 0 \), Fig.~\ref{fig:Figure7}(b) and Figs.~\ref{fig:Figure7}(g); quadratic algebraic evolution at the EP3, Fig.~\ref{fig:Figure7}(c) and Figs.~\ref{fig:Figure7}(h); and linear Fig.~\ref{fig:Figure7}(d) and Figs.~\ref{fig:Figure7}(i), and hyperbolic, Fig.~\ref{fig:Figure7}(e) and Fig.\ref{fig:Figure7}(j), evolution at the two EP2s.
Detailed analysis shows that the initial state is not recovered, nor transferred, but propagation is highly correlated. 

\begin{figure}[h]
    \begin{center}
        \includegraphics[scale = 1]{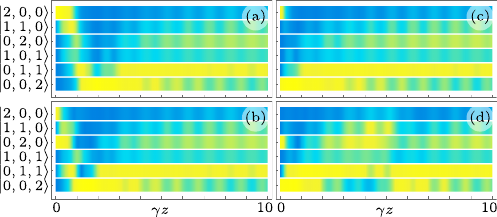}
        \caption{Propagation-dependent dynamics for three clockwise turns along a circular loop in the $(\kappa_{1},\kappa_{2})$ plane centered at EP3 with initial state (a) $ \vert 2,0,0 \rangle $ and NOON states
        (b) $( \vert 2,0,0 \rangle + \vert 0 ,2 ,0 \rangle) / \sqrt{2}$
        (c) $( \vert 2,0,0 \rangle + \vert 0 ,0 ,2 \rangle) / \sqrt{2}$
        (d) $( \vert 0,2,0 \rangle + \vert 0 ,0 ,2 \rangle) /\sqrt{2}$.} 
        \label{fig:Figure8}
    \end{center}
\end{figure}

Figure~\ref{fig:Figure8} shows the renormalized amplitudes $P_{n_{1},n_{2},n_{3}}(z)$ for two-excitation inputs under propagation along three clockwise loops around EP3 in the $(\kappa_{1},\kappa_{2})$ plane.
We explore an initial Fock state $\vert 2,0,0\rangle$, Fig.~\ref{fig:Figure8}(a), 
and three NOON states: $(\vert 2,0,0\rangle+\vert 0,2,0\rangle)/\sqrt{2}$ in Fig.~\ref{fig:Figure8}(b), 
$(\vert 2,0,0\rangle+\vert 0,0,2\rangle)/\sqrt{2}$ in Fig.~\ref{fig:Figure8}(c), 
and $(\vert 0,2,0\rangle+\vert 0,0,2\rangle)/\sqrt{2}$ in Fig.~\ref{fig:Figure8}(d).
The loop dynamics amplifies and mixes components, producing correlated propagation that depends on the input superposition. 

\begin{figure}[h]
    \begin{center}
        \includegraphics[scale = 1]{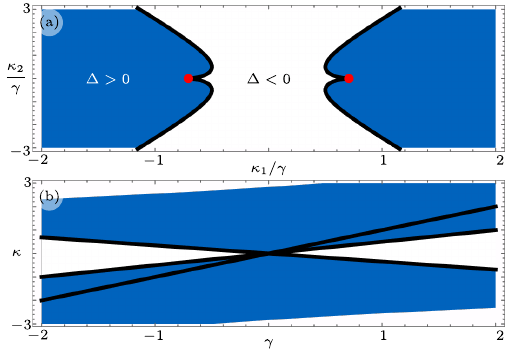}
        \caption{Sign of the discriminant with $\Delta>0$ in blue, $\Delta<0$ in white, $\Delta=0$ in black for EP2, and red for EP3 for the non-Hermitian chiral families in (a) Eq. (\ref{eq:Chi1}) and (b) Eq. (\ref{eq:Chi2}).}
        \label{fig:Figure9}  
    \end{center}
\end{figure}

We close by noting that our algebraic approach makes it straightforward to identify exceptional points.
Consider the chiral family,
\begin{align}
\bm{M}_{1} =
\begin{pmatrix}
i \gamma & \kappa_{1} & i \kappa_{2} \\
\kappa_{1} & 0 & \kappa_{1} \\
i \kappa_{2} & \kappa_{1} & -i \gamma
\end{pmatrix}, \label{eq:Chi1}
\end{align} 
that supports the same EP3 as our example but the EP2 line is deformed by the chirality, Fig.~\ref{fig:Figure9}(a). 
In contrast, another chiral cyclic family,
\begin{align}
\bm{M}_{1} =
\begin{pmatrix}
i \gamma & \kappa & -\kappa \\
\kappa & 0 & i \kappa \\
-\kappa & i \kappa & -i \gamma
\end{pmatrix}, \label{eq:Chi2}
\end{align}
has no EP3 beyond the trivial matrix and shows only EP2 lines, see Fig.~\ref{fig:Figure9}(b).
In both cases the discriminant is real.

\section{Conclusion} \label{sec:conclusions}
We developed a symmetry-guided framework for non-Hermitian $N$-mode couplers starting from $\mathrm{gl}(N,\mathbb{C})$, the most general Lie algebra of complex $N \times N$ matrices, which includes arbitrary non-Hermitian mode-coupling matrices without additional assumptions.
A gauge renormalization removes global phase and uniform gain or loss.
In the classical regime, our approach yields complex traceless matrices in $\mathrm{sl}(N,\mathbb{C})$. 
In the quantum regime, it yields a bosonic embedding that preserves the total excitation number and realizes irreducible representations of $\mathrm{sl}(N,\mathbb{C})$.
Our framework determines spectral properties through trace invariants and discriminants and yields an exact propagator via a Wei--Norman factorization at the level of the full algebraic class. 
It avoids case-by-case analysis and identifies families of couplers as symmetry-restricted submanifolds of $\mathrm{sl}(N,\mathbb{C})$.

We developed the three-mode coupler in the isospin--hypercharge representation of $ \mathrm{sl}(3,\mathbb{C}) $, where the Cartan generators describe relative phases and differential gains, and the ladder generators represent coupling between pairs of modes in the physical system.
In the classical regime, it corresponds to traceless non-Hermitian $3 \times 3$ matrices that map directly onto the coupler, with each component of the state vector representing the field amplitude in the corresponding mode.
In the quantum regime, it corresponds to a three-boson embedding that restricts dynamics to fixed-$n$ Fock subspaces.
We describe the state vector in the totally symmetric irreducible representation $(n,0)$, with triangular weight diagrams in the isospin-hypercharge plane.
The single excitation sector reproduces the classical model, as expected, while higher excitations resemble multiplets in hadronic physics. 
This bosonic embedding provides synthetic dimensions, where higher-dimensional algebraic dynamics unfold in a compact three-mode system, and conversely guides the design of classical systems that simulate the quantum dynamics.

We studied the spectral structure of the three-mode coupler within the $\mathrm{SL}(3,\mathbb{C})$ group using a normal-ordered similarity transformation.
In the classical regime, we build the local spectrum from a gauge that removes the Cartan sector and connects the problem to the characteristic polynomial, a depressed cubic defined by trace invariants. 
A propagation-dependent gauge in the Cartan sector enforces coincidence of the local and dynamical spectra, reveals the geometric phase as a holonomy, and distinguishes adiabatic from exact propagation.
In the quantum regime, we extend the classical construction to the bosonic bilinear form embedding, providing the corresponding spectral structure without altering the underlying algebraic framework.
Our Lie group approach makes explicit and tractable the role of local and dynamical gauges in non-Hermitian dynamics in both regimes, a perspective that, to our knowledge, is lacking in the field.

We constructed an explicit propagator for the three-mode coupler using a normal-ordered Wei-Norman decomposition.
In the classical regime, this yields a triangular hierarchy of differential equations: a coupled Riccati pair, a scalar Riccati equation, and linear relations for the remaining variables.
They can be solved sequentially under analytic conditions on the coupler propagation-dependent parameters.
In the quantum regime, the same classical envelopes drive the propagation after promoting the matrices to their bosonic bilinear form. 
Our approach traces orbits on the group manifold, with the spectrum fixing the tangent and integrations generating the full orbit, turning spectral structure into exact state evolution.
These orbits may be compact, corresponding to bounded oscillation, or non-compact, corresponding to unbounded polynomial or hyperbolic propagation.

Recent experiments demonstrated the interplay between EPs and the non-Hermitian skin effect (NHSE) in coupled photonic lattices with artificial gauge fields \cite{Wang2025p339}. 
Their observations that EPs can compress the eigenvalue spectrum and modulate or suppress skin localization highlight the role of EPs as a controllable degree of freedom in non-Hermitian systems. 
In addition, these results provide an experimental foundation that closely parallels and supports the theoretical framework developed in our three-mode coupler model. 
This connection not only reinforces the universality of EP–NHSE interplay but also opens new possibilities for engineered control of non-Hermitian dynamics in multi-mode quantum and photonic platforms.


\section*{Funding}
National Science Foundation (PHY-2012172, OSI-2231387); Bayerische Staatsministerium für Wirtschaft, Landesentwicklung und Energie (6GQT); Bundesministerium für Bildung und Forschung (16KISK002, 16KIS1598K, 16KISQ039, 16KISQ077, 16KISQ093); Deutsche Forschungsgemeinschaft (1129/2-1).

\section*{Acknowledgments}
B.~M.~R.~L. acknowledges support and hospitality as an affiliate visiting colleague at the Department of Physics and Astronomy, the University of New Mexico.
This work was supported by NSF grants PHY-2012172 and OSI-2231387.\\
This research is part of the Munich Quantum Valley, which is supported by the Bavarian state government with funds from the Hightech Agenda Bayern Plus and received support from the Bavarian Ministry for Economic Affairs (StMWi) via the project 6GQT.
The authors acknowledge the financial support by the Federal Ministry of Education and Research of Germany in the programme of “Souverän. Digital. Vernetzt.” for the Joint project 6G-life, project identification number: 16KISK002 and via grants 16KIS1598K, 16KISQ039, 16KISQ077, 16KISQ093 and that of the DFG via grant 1129/2-1.

\section*{Disclosures}
The authors declare no conflicts of interest.

\section*{Data Availability Statement}
All the data is available from the corresponding author upon reasonable request.


\bibliography{references}

\end{document}